# A Nonparametric Reconstruction of the Hubble Parameter $H(z)$ Based on Radial Basis Function Neural Networks

Jian-Chen Zhang[1,8], Yu Hu[2,3,8], Kang Jiao[4,5], Hong-Feng Wang[2,3], Yuan-Bo Xie[5,6], Bo Yu[7], Li-Li Zhao[2,3], and Tong-Jie Zhang (张同杰)[3,5,6]

[1] Shandong College of Electronic Technology, Jinan, 250200, People's Republic of China
[2] College of Computer and Information Engineering, Dezhou University, Dezhou 253023, People's Republic of China
[3] Institute for Astronomical Science, Dezhou University, Dezhou 253023, People's Republic of China; tjzhang@bnu.edu.cn
[4] School of Physics and Microelectronics, Zhengzhou University, Zhengzhou 450001, People's Republic of China
[5] Department of Astronomy, Beijing Normal University, Beijing 100875, People's Republic of China
[6] Institute for Frontiers in Astronomy and Astrophysics, Beijing Normal University, Beijing 102206, People's Republic of China
[7] School of Mathematics and Big Data, Dezhou University, Dezhou 253023, People's Republic of China

## Abstract

Accurately measuring the Hubble parameter is vital for understanding the expansion history and properties of the Universe. In this paper, we propose a new method that supplements the covariance between redshift pairs to improve the reconstruction of the Hubble parameter using the observational Hubble data set. Our approach uses a cosmological model-independent radial basis function neural network to effectively describe the Hubble parameter as a function of redshift. Our experiments show that this method results in a reconstructed Hubble parameter of $H_0 = 67.1 \pm 9.7 \, \text{km} \, \text{s}^{-1} \, \text{Mpc}^{-1}$, which is more noise resistant and fits the $\Lambda$CDM model at high redshifts better. Providing the covariance between redshift pairs in subsequent observations will significantly improve the reliability and accuracy of Hubble parametric data reconstruction. Future applications of this method could help overcome the limitations of previous methods and lead to new advances in our understanding of the Universe.

*Unified Astronomy Thesaurus concepts:* Observational cosmology (1146); Computational methods (1965); Hubble constant (758); Cosmological parameters (339)

## 1. Introduction

Measuring the Hubble parameter $H(z)$ is a fundamental task in cosmology, as it allows us to understand the current expansion rate of the Universe and the history of its evolution (Moresco et al. 2022). However, it is callenging to accurately determine $H_0$ due to a variety of factors, including distance measurements and observational uncertainties. In recent years, the so-called "Hubble tension" (Verde et al. 2019; Freedman 2021) has emerged as an intriguing puzzle in cosmology, referring to the discrepancy between the local measurement of $H_0$ based on the cosmic distance ladder (Riess et al. 2016) and the value inferred from observations of the cosmic microwave background radiation (Planck Collaboration et al. 2020). This tension suggests the possibility of new physics beyond the standard cosmological model or systematic errors in the data analysis (Di Valentino et al. 2021). Nonparametric techniques based on observational data sets have been developed as an alternative method to address these challenges in a model-independent way.

One of the most popular nonparametric methods for reconstructing $H_0$ is the Gaussian process (GP) proposed by Williams & Rasmussen (1995), which has been successfully used to reconstruct $H_0$ using various data sets by Gómez-Valent & Amendola (2018), as well as the teleparallel gravity Lagrangian by Said et al. (2021). Sun et al. (2021) investigated the influence of hyperparameters within the GP on the reconstruction of the Hubble constant, demonstrating that considering lower and upper bounds on the GP hyperparameters is necessary to reliably and robustly extrapolate results. The GP assumes that each element of the data set is normally distributed as part of a larger random process. By optimizing the covariance function between these points, the entire evolutionary process of the data set can be reconstructed for certain data. However, the GP method faces several challenges in the context of reconstructing $H_0$ from observational data. One problem is that not all cosmological data follow a normal distribution, which affects the reliability of the reconstructed $H(z)$. In addition, the choice of the covariance function is an auxiliary problem, although this can be partially avoided by using the genetic algorithms in Bernardo & Said (2021). Another challenge is that the GP method can overfit to low-redshift data, according to Colgáin & Sheikh-Jabbari (2021), which affects the inferred value of $H_0$ in the reconstructed Hubble parameter function. Addressing these challenges is crucial for improving the accuracy and reliability of Hubble parameter measurements using nonparametric methods. We mitigate the risk of overfitting in our approach through the implementation of ridge regularization and cross validation techniques. Another nonparametric method used to reconstruct Hubble parameter is the principal component analysis (PCA) introduced by Ishida & de Souza (2011) and further developed by Sharma et al. (2022). Although Sharma et al. (2022) combined the PCA algorithm and calculation of correlation coefficients, the PCA algorithm reduces dimensionality by focusing on the directions of the highest variance, potentially discarding information present in the less dominant components. This could result in a loss of details that may be crucial for an accurate reconstruction. It also assumes that the underlying structure is linear. If the relations or variances in the data are nonlinear, PCA might not capture these complexities effectively. Additional nonparametric techniques in this field encompass the locally weighted scatterplot smoothing (LOWESS) introduced by Montiel et al. (2014) and the charm code introduced by Porqueres et al. (2017). The LOWESS method uses a combination of the Loess method and the Simex method, and it aims to reconstruct cosmic expansion without any predefined cosmological models. While combining two methods can capitalize on their strengths, it can also compound their weaknesses. There might be scenarios where the assumptions or limitations of one method negatively impact the overall performance of the combined approach. In addition, the Loess component relies on a smoothing parameter that dictates the flexibility of the regression function. Choosing an inappropriate value for this parameter can lead to overfitting or underfitting. The charm code formulates the problem of recovering the signal from data in a Bayesian manner. Even though the charm method does not assume an analytical form for the density evolution, it still assumes its smoothness with the scale factor. Any deviation from this assumption might impact the accuracy of its results. Moreover, being fully Bayesian means that the charm method might be sensitive to the choice of prior. A poor choice of prior can lead to biased results.

---

[8] These authors contributed equally to this work.

The concept of artificial neural networks (ANNs) was first proposed by McCulloch & Pitts (1943). ANNs are a type of nonparametric neural network method designed to fit nonlinear functions. Inspired by biological neural networks, the structure of ANNs comprises several layers of neurons, each with an activation function that regulates the output of the neuron. Therefore, each neuron contains a certain number of hyperparameters that must be assigned. To do this, the network is trained using observed data that optimize the neural response, resulting in a neural network that can simulate the observed data at each redshift. Compared to GP, which often suffer from overfitting and kernel selection issues, ANNs can effectively remedy these issues by optimizing a large number of hyperparameters. This means that compared to GP, the trained ANN better models the natural process that produces the observed results, and with a sufficient number of hidden nodes, it can approximate any nonlinear function with arbitrary precision. Dialektopoulos et al. (2022) introduced another advantage of the ANN method, which is that its $H(z)$ reconstruction is largely independent of priors, while GP is significantly affected by different priors. These fundamental distinctions arise because the ANN structure makes fewer assumptions than GP in this context, providing a more authentic representation of cosmological parameters. ANNs are typically designed as nonprobabilistic models, which means that they directly approximate functions without a predefined probabilistic structure. Their flexibility arises from their architecture, particularly from the number of layers and neurons, and the nonlinear activations. As such, while training data and the objective function guide the learning process, ANNs do not inherently rely on prior beliefs about the data distribution. On the other hand, GPs are rooted in Bayesian inference, where prior beliefs about the function space play a central role. The GP prior, typically defined by a mean function and a covariance kernel, represents our preliminary beliefs about the function before observing any data. Upon observing data, the GP posterior is computed by updating this prior, using the Bayes theorem. Consequently, the choice of prior can significantly influence the GP behavior and predictions, making it more sensitive to the prior selection than ANNs.

Numerous scholars have already used ANN-based machine-learning models to investigate the Hubble parameters and related research. Wang et al. (2021) proposed a new method of likelihood-free cosmological inference that uses two ANN models, MAF and denoising autoencoders (DAE), to accurately estimate posterior distributions and extract features from observational data without relying on tractable likelihoods, demonstrating its efficacy on simulated and real data. Chen et al. (2023) explored the effectiveness of information-maximizing neural networks (IMNN) and DAE in compressing observational Hubble data (OHD) for cosmological parameter estimation, and found that IMNN provides better dimensionality reduction and robustness than DAE. Wang et al. (2020) proposed a nonparametric approach for reconstructing functions from observational data using an ANN and tested it on Hubble parameter measurements and the distance-redshift relation of Type Ia supernovae. However, it does not take covariance between redshift points into account.

The radial basis function neural network (RBFNN), introduced by Broomhead & Lowe (1988), is a special type of neural network in ANN. While ANNs use activation functions as bases to overlay and fit the target function, the RBFNN uses special bases such as Gaussian kernel functions to overlay and fit the target function. Compared to ANNs, the RBFNN has a strong generalization ability, optimal universal function approximation ability, powerful noise resistance, and faster convergence speed. We use the RBFNN as a new regression method for reconstructing the Hubble parameter, and it shows good fitting and prediction performance.

## 2. Method

In this section, we discuss the RBFNN derived from an ANN. We begin by introducing the features and development history of ANNs. Then, we explain how an RBFNN is used in our work. Finally, in Section 2.3, we cover the learning algorithm for RBFNN.

### 2.1. Artificial Neural Network

An ANN is a digital network structure that imitates the characteristics of a biological neural network and connects it with simple processing units. These networks use a unique mechanism, learning, to obtain information about the processing system and store this information in the network structure and parameters. When dealing with a system with a neural network, only the input and output of the system are generally of concern, and the internal process is regarded as a black box. This approach allows for a simplified understanding of the system by ignoring its complex internal relations. An ANN can be considered an intelligent tool that reduces the computation workload by providing imprecise solutions, corresponding to the precise computations in computers starting in the last century. This reduced computation workload is important for many applications.

Many practical systems have complex internal structures that can be difficult to understand and model. In contrast, neural networks have simple structures that are easy to work with and analyze. Furthermore, their structures are conducive to parallel processing, which can significantly speed up computations. Neural networks also demonstrate better fault tolerance and the ability to learn adaptively, making them well suited for a wide range of applications.

Since the 1980s, ANNs have undergone extensive research and development, leading to significant breakthroughs in areas such as image processing, pattern recognition, fault diagnosis, signal processing, and automatic control. Today, neural networks are widely used and play a critical and irreplaceable role in many fields.

Scholars have proposed various types of networks to address different application requirements. For example, the backpropagation network introduced in Hecht-Nielsen (1992) is widely used for supervised learning tasks, while the Hopfield network introduced by Hopfield (1982) is commonly used for associative memory applications. The wavelet network combines wavelet analysis and neural network techniques for signal processing, and the radial basis function (RBF) network is used for function approximation and classification tasks. Other types of networks include adaptive networks, which are used for reinforcement learning, and many others. Each of these networks has unique characteristics and is used in specific applications to achieve optimal performance.

### 2.2. Radial Basis Function Neural Network

The RBFNN is a crucial type of neural network in ANN methodologies (Broomhead & Lowe 1988). RBFNNs are known for their simple structure, fast training speed, and nonlinear processing ability. They are commonly used in various applications, including face recognition by Er et al. (2002), signal processing by Dhanalakshmi et al. (2009), and function approximation by Huang et al. (2005). In this paper, we use the RBFNN to analyze Hubble parameter observational data obtained from the cosmic chronometer method.

The data consist of 32 points, each representing a redshift value. The RBFNN used in this analysis has a $1 \times 32$ input layer, 10,000 hidden layer nodes, and a two-node output layer. The output layer consists of $y_1$, which represents $H(z)$, and $y_2$, which represents $\sigma(z)$. The topology of the RBFNN used in this paper is shown in the figure below. We aim to use the RBFNN to analyze the data and gain insights into the behavior of the Universe at different redshifts.

The RBF network is composed of two main layers: the nonlinear transformation layer, and the linear merging layer. The nonlinear transformation layer applies an RBF to the input data, which maps the input space into a higher-dimensional feature space. This transformation allows the network to model complex nonlinear relations in the input data. In Figure 1, the RBF network is assumed to have $N$ inputs, which are transformed by $K$ hidden neurons in the nonlinear transformation layer. The output of the nonlinear transformation layer is then fed into the linear merging layer, where it is combined to produce $P$ outputs. The purpose of the RBF network is to learn a mapping between the input data and the desired output. The output of the linear transform layer is

$$h_i = \phi(\boldsymbol{x}) = \phi\left(\frac{\|\boldsymbol{x} - \boldsymbol{c}_i\|}{\sigma_i}\right), \ i = 1, 2, \cdots, K. \quad (1)$$

Here, $\boldsymbol{x} = \{x_i | i = 1, 2, \cdots N\}$ is the input vector, $\boldsymbol{c}_i = \{c_i(k) | k = 1, 2, \cdots N\}$ is the center vector of the $i$th nonlinear transformation unit, $c_i(k)$ represents the $k$th component of the $i$th center, $\sigma_i$ is the width of the $i$th nonlinear

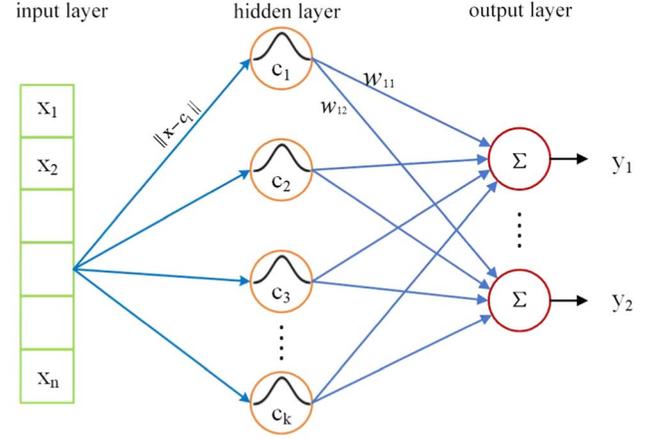

**Figure 1.** Structure of the RBFNN.

transformation unit, $\| * \|$ represents the norm, usually taking the two-norm, $\phi(\cdot)$ represents the nonlinear output function. From the properties of the norm, we know that it has radial symmetry, so $\phi(\cdot)$ is called the RBF. The RBF is generally called a Gaussian function, that is,

$$\phi(x) = \exp\left(-\frac{x^2}{2\sigma^2}\right). \quad (2)$$

The second layer is the linear merging layer, which converts $K$ hidden neurons into $P$ outputs. The role of this layer is to combine the outputs of the transform layer linearly weighted,

$$y_i = \sum_{j=1}^{K} h_j w_{ji}, \ (1 \leqslant i \leqslant P) + w_0. \quad (3)$$

$w_0$ represents the bias neuron, and $w_{ji}$ is the connection weight from the hidden layer to the output layer. If the weights are rewritten in vector form and let

$$\boldsymbol{\Phi} = \begin{bmatrix} \phi(\|x_1 - c_1\|) & \phi(\|x_1 - c_2\|) & \phi(\|x_1 - c_N\|) \\ \phi(\|x_2 - c_1\|) & \phi(\|x_2 - c_2\|) & \phi(\|x_2 - c_N\|) \\ \phi(\|x_N - c_1\|) & \phi(\|x_N - c_2\|) & \phi(\|x_N - c_N\|) \end{bmatrix}, \quad (4)$$

then Equation (3) can be written in matrix form,

$$\boldsymbol{y} = \boldsymbol{w}\boldsymbol{\Phi} + \boldsymbol{w}_0, \quad (5)$$

where $\boldsymbol{\Phi}$ is called the RBF matrix.

Generally speaking, the application of the RBF neural network includes two processes, namely, constructing and learning. To construct an RBFNN, we first determine the neural network input and target set. According to the characteristics of the input vector, we determine the number of neurons in the hidden layer and the center of the RBF, and we again estimate the width of the RBF. The derivation of the center and width parameters can be approached using clustering techniques. However, an alternative method considers every data point as a potential center and employs ridge regularization to stave off overfitting. In this work, we adopt the latter method. Ridge regression is an adept regularization strategy widely used to counteract the overfitting phenomenon in various machine-learning models, including the comprehensive RBFNN. The essence of ridge regression lies in its ability to strike a balance: it aims to minimize the model training-data error while concurrently curbing the magnitude of its parameters. This dual objective compels the model to embrace simplicity, preventing it from being swayed by noise and subsequently boosting its generalization capability. When applied to a full RBFNN, ridge regression regulates the weights corresponding to the RBFs and output neurons. This regulation mitigates the influence of any single weight, ensuring that no singular parameter overly dictates the learning trajectory. As a result, the RBFNN becomes inherently resilient against overfitting, proficiently capturing the core trends in the data set and sidestepping extraneous intricacies.

Finally, the initial value of the weight is given, and the learning phase activates. The learning process of the RBFNN is iterative; that is, some algorithms (e.g., gradient descent) are used to determine the iterative direction and gradually adjust the connection weights from the hidden layer to the output layer, the center, and the width of the RBF. In our study, we implement the early stopping strategy of limiting the number of iteration steps to a fixed value of 100 as one of the two available approaches. The other approach is a fixed error threshold.

### 2.3. RBFNN Learning Algorithm

Let $\mathbf{y}'$ be the output of the network and $\mathbf{y}$ the objective function of the network. We then define the error function as

$$E = \frac{1}{P}\sum_{i=1}^{P} \|\mathbf{y}'_i - \mathbf{y}_i\|^2. \tag{6}$$

It can be seen that $E$ is a function of $\mathbf{c}$, $\mathbf{w}$, and $\sigma$. The RBF network training process adjusts the above three sets of parameters so that $E$ tends to be the smallest. The primary method of feed-forward neural network learning is reverse error-propagation, and the commonly used error-propagation algorithm is the gradient descent algorithm,

$$\mathbf{c}^{\text{new}} = \mathbf{c}^{\text{old}} - r\Delta\mathbf{c} \tag{7}$$

$$\mathbf{w}^{\text{new}} = \mathbf{w}^{\text{old}} - r\Delta\mathbf{w} \tag{8}$$

$$\sigma^{\text{new}} = \sigma^{\text{old}} - r\Delta\sigma. \tag{9}$$

The value of $r$ is between 0 and 1, called the learning rate, which is used to adjust the learning speed and avoid the oscillation phenomenon of learning. Generally, the learning process is to iterate the above three equations alternately.

## 3. Experiment

In this section, we introduce a method based on the simulated $H(z)$ using the Hubble parameter observational data. The RBFNN model is optimized through the simulated $H(z)$ data, and then the best RBFNN model is used to obtain the Hubble parameter reconstruction data based on the Hubble parameter observational data.

### 3.1. Generating Simulated Data

To optimize the neural network model used for the reconstruction of Hubble parameter observational data, Wang et al. (2020) simulated $H(z)$ using a flat $\Lambda$CDM model with a fiducial Hubble constant of $H_0 = 70 \, \text{km s}^{-1} \, \text{Mpc}^{-1}$ and a matter energy density parameter of $\Omega_m = 0.3$. This paper uses the same approach to simulate $H(z)$ used in the RBFNN, which is

$$H(z) = H_0\sqrt{\Omega_m(1+z)^3 + 1 - \Omega_m}, \tag{10}$$

with the fiducial $H_0 = 70 \, \text{km s}^{-1} \, \text{Mpc}^{-1}$ and $\Omega_m = 0.3$.

**Table 1**
Measurements of CC $H(z)$ Using the Differential Age Method

| $z$ | $H(z)$ (km s$^{-1}$ Mpc$^{-1}$) | References |
|---|---|---|
| 0.09 | 69 ± 12 | Jimenez et al. (2003) |
| 0.17 | 83 ± 8 | |
| 0.27 | 77 ± 14 | |
| 0.4 | 95 ± 17 | |
| 0.9 | 117 ± 23 | Simon et al. (2005) |
| 1.3 | 168 ± 17 | |
| 1.43 | 177 ± 18 | |
| 1.53 | 140 ± 14 | |
| 1.75 | 202 ± 40 | |
| 0.48 | 97 ± 62 | Stern et al. (2010) |
| 0.88 | 90 ± 40 | |
| 0.1791 | 75 ± 4 | |
| 0.1993 | 75 ± 5 | |
| 0.3519 | 83 ± 14 | |
| 0.5929 | 104 ± 13 | Moresco et al. (2012) |
| 0.6797 | 92 ± 8 | |
| 0.7812 | 105 ± 12 | |
| 0.8754 | 125 ± 17 | |
| 1.037 | 154 ± 20 | |
| 0.07 | 69 ± 19.6 | |
| 0.12 | 68.6 ± 26.2 | Zhang et al. (2014) |
| 0.2 | 72.9 ± 29.6 | |
| 0.28 | 88.8 ± 36.6 | |
| 1.363 | 160 ± 33.6 | Moresco (2015) |
| 1.965 | 186.5 ± 50.4 | |
| 0.3802 | 83 ± 13.5 | |
| 0.4004 | 77 ± 10.2 | |
| 0.4247 | 87.1 ± 11.2 | Moresco et al. (2016) |
| 0.4497 | 92.8 ± 12.9 | |
| 0.4783 | 80.9 ± 9 | |
| 0.47 | 89 ± 49.6 | Ratsimbazafy et al. (2017) |
| 0.80 | 113.1 ± 28.2 | Jiao et al. (2023) |

The redshift of the observable $H(z)$ (Table 1) is assumed to be susceptible to a Gamma distribution,

$$p(x; \alpha, \lambda) = \frac{\lambda^\alpha}{\Gamma(\alpha)}x^{\alpha-1}e^{-\lambda x}, \tag{11}$$

where $\alpha$ and $\lambda$ are parameters, and the gamma function is

$$\Gamma(\alpha) = \int_0^\infty e^{-t}t^{\alpha-1}dt. \tag{12}$$

Figure 2 illustrates the distribution of the observed $H(z)$ and the anticipated distribution function for the redshift $z$.

We display the errors against redshift $z$ in Figure 2. The error in $H(z)$ clearly grows with increasing redshift. Following Ma & Zhang (2011), we assume that the error in $H(z)$ rises linearly with redshift. We begin by fitting $H(z)$ with polynomials of the first degree, obtaining $\sigma_0 = 9.72z + 14.87$ (dashed blue line). In this case, we assume that $\sigma_0$ represents the mean value of $H(z)$ at a given redshift. Then, two lines (solid green lines) are picked symmetrically around the mean value line to guarantee that the majority of data points fall inside their region, and these two lines have functions $\sigma_- = 2.92z + 4.46$ and $\sigma_+ = 16.52z + 25.28$. Finally, the error $\hat{\sigma}(z)$ is created randomly using the Gaussian distribution $\mathcal{N}(\sigma_0(z), \varepsilon(z))$, with $\varepsilon(z) = (\sigma_+ - \sigma_-)/4$ chosen to guarantee that $s(z)$ falls inside the region with a 95% probability. When using Equation (11) to produce fiducial values of the Hubble parameter $H_{\text{fid}}(z)$, the fiducial values are simulated randomly by adding $\Delta H$ subject to $\mathcal{N}(0, \hat{\sigma}(z))$. Thus, the ultimate simulated Hubble parameter is $H_{\text{sim}}(z) = H_{\text{fid}}(z) + \Delta H$, where the uncertainty is an unknown quantity $\hat{\sigma}(z)$. As a result, one may replicate samples of the Hubble parameter in the flat $\Lambda$CDM model with the expected redshift and error distributions. We highlight that the mock $H(z)$ is used to improve the network model and that the assumption that the $H(z)$ error grows linearly with redshift has no effect on the reconstruction of the observational $H(z)$ (Ma & Zhang 2011); consequently, the $H(z)$ error model is acceptable in our study.

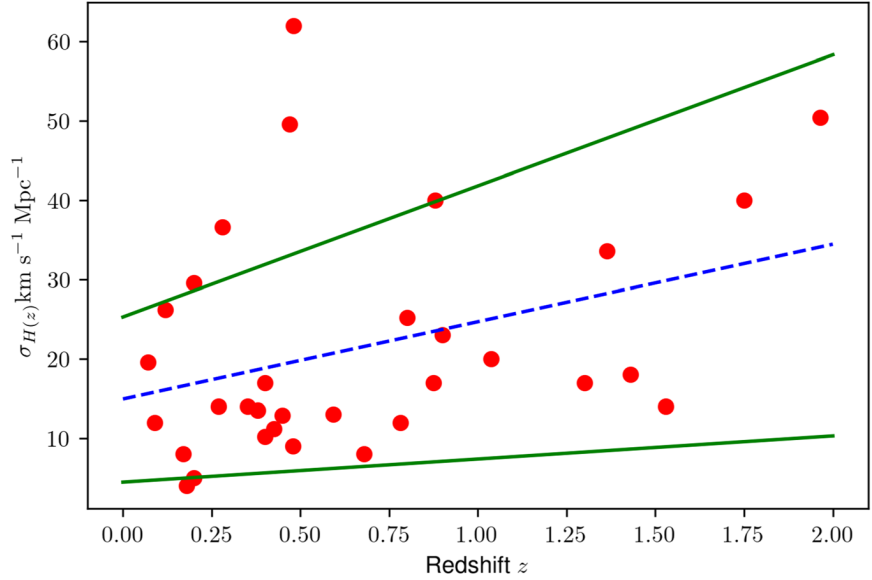

**Figure 2.** Reconstructed errors of the observational $H(z)$ with linear regression. The solid red dots represent the observational error of $H(z)$. The heuristic bounds $\sigma_+$ and $\sigma_-$ are plotted as the two solid green lines. The dashed line shows the estimated mean uncertainty $\sigma_z$.

### 3.2. RBFNN Model Training

In this section, we illustrate how to train the RBFNN using mock data $H(z)$ gained in Section 3.1. The data required to train the network are simulated using the flat $\Lambda$CDM model and the approach described in Section 3.1. The simulation Hubble parameter has 32 data points, which is the same number as for the observational data.

The hypothesis of an RBFNN is as follows:

$$h(x) = \text{Output}\left(\sum_{m=1}^{M} w_{ij}\text{RBF}(x, c_i) + w_0\right), \quad (13)$$

where $c_i$ is called the center, and $w_{ij}$ is the voting parameter.

Our RBFNN model uses a Gaussian RBF function, which is commonly used in RBFNNs. The function is defined as $\text{RBF}(\boldsymbol{x}_n, \boldsymbol{x}_i) = \exp(-\gamma\|\boldsymbol{x}_n - \boldsymbol{x}_i\|^2)$, where $\gamma$ is a hyperparameter that controls the width of the RBF function.

We use a full RBFNN model, where the first key variable $c_i$ is determined by all 32 data points. The second key variable $w_{ij}$ is computed using ridge regularization, which helps to prevent overfitting. Specifically, we set $w_{ij} = (Z^T Z + \lambda I)^{-1} Z^T \boldsymbol{y}$, where $z_n$ is a vector containing the RBF values between the $n$th data point and all other data points, $Z$ is a matrix containing all $z_n$ vectors, $\boldsymbol{y}$ is the vector of target outputs, and $\lambda$ is a regularization parameter that controls the strength of the regularization.

By using ridge regularization, we avoid the potential overfitting problem that can arise when the error of the RBFNN model is zero. This occurs because the model may fit the training data too closely, leading to a poor generalization performance on new data. Ridge regularization adds a penalty term to the objective function of the model that discourages overfitting by shrinking the magnitude of the weight parameters of the model. This results in a model that has a better balance between fitting the training data and generalizing to new data.

### 3.3. Hyperparameter Tuning

The optimization of hyperparameters is an important step in building accurate and robust machine-learning models. Hyperparameters are the parameters of a machine-learning model that are not learned from the data, but are set by the user. Examples of hyperparameters include the learning rate, regularization parameter, and number of hidden layers in a neural network. Cross validation and grid search are common techniques that are used to optimize hyperparameters. In this study, we applied cross validation and grid search to optimize the hyperparameters of a machine-learning model using a mock OHD data set.

The grid search (GS) is an approach that exhaustively constructs estimators with all the hyperparameter combinations. LaValle et al. (2004) have proved that it is an effective way for hyperparameter turning, while cross validation is a method that is less biased or has a less optimistic estimate of the model performance compared to other techniques, such as a straightforward train/test split.

We used a data set consisting of 32 data points and divided it into five separate groups using random shuffling. For each distinct group, we designated the group as the test data set and used the remaining groups as the training data set. We trained a model with different combinations of hyperparameters using GS on the training data set and assessed its performance on the test data set using the mean square error (MSE) metric. We tested a range of hyperparameters, including learning rates $r$ of 0.001, 0.01, and 0.1, the values of the regularization term $\lambda$ ranging from 0 to 1 with a step size of 0.1, and the number of neuron values ranging in an array [128, 256, 1000, 3000, 4096, $10^4$, $10^5$]. We selected the hyperparameters with the best performance based on the average MSE in all five groups. We found that the optimal hyperparameters were a learning rate of 0.01, an RBF parameter $\gamma$ of 0.2, a regularization parameter $\lambda$ of 0.3, and a number of neurons of $10^5$.

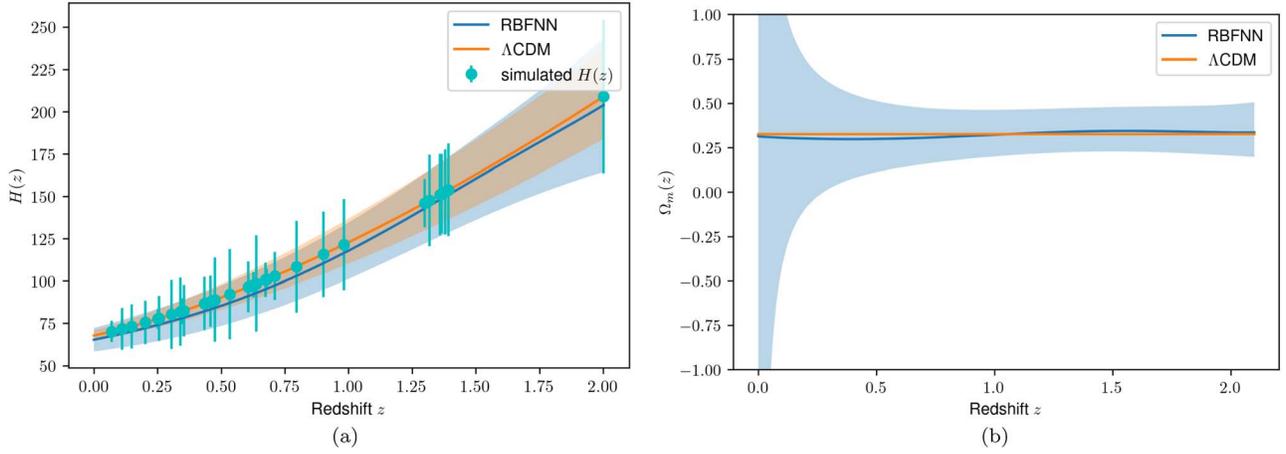

**Figure 3.** The reconstructed functions of mock $H(z)$ (solid blue curve) and its corresponding $1\sigma$ error with RBF networks (a). The green dots with the error bars represent the simulated $H(z)$ data and the corresponding $1\sigma$ error, while the orange curve corresponds to the fiducial flat $\Lambda$CDM model with $H_0 = 67.824$ km s$^{-1}$ Mpc$^{-1}$ and $\Omega_m = 0.326$. In the left panel, the shaded regions illustrate the reconstruction errors associated with both the $\Lambda$CDM model and the RBFNN model when using simulated $H(z)$ data, while the right panel portrays the propagation error.

### 3.4. Results of the Mock Data Reconstruction

After a GS cross validation, the optimal RBFNN model was obtained. Using this optimal RBFNN, the simulated Hubble parameter values from the previous section were reconstructed, and the result is shown in Figure 3(a). In order to test the RBFNN model, we refer to Zunckel & Clarkson (2008), Sahni et al. (2008), and Shafieloo & Clarkson (2010) to carry out further calculations for $\Omega_m$,

$$\Omega_m^{(1)}(z) = \frac{h^2 - 1}{z(3 + 3z + z^2)}, \quad (14)$$

where $h$ is $H(z)/H(0)$ predicted by the RBF model we reconstruct, as shown in Figure 3(b).

This figure shows that the reconstructed function agrees well with the flat $\Lambda$CDM model with a prior $H_0 = 67.824$ km s$^{-1}$ Mpc$^{-1}$, and the $\Omega_m$ curve hovers around the flat $\Lambda$CDM model with prior $\Omega_m = 0.326$, indicating the reliability of the model we construct.

### 3.5. Predicting H(z)

In this section, based on the hyperparameters obtained in the previous subsection, the model is trained using the Hubble parameter observational data, and the reconstruction function of $H(z)$ is visualized. First, a complete RBFNN is constructed with 32 centers, which is the same as the number of observational data. In consideration of the possibility of overfitting, a regularization term is added to the model to avoid this issue by limiting the number of centers and voting weights.

The reconstruction results show that $H_0 = 67.072^{+9.7}_{-9.7}$ km s$^{-1}$ Mpc$^{-1}$. Figure 4 compares the constructed RBFNN with the flat $\Lambda$CDM model with prior $H_0 = 67.824$ km s$^{-1}$ Mpc$^{-1}$ and $\Omega_m = 0.326$. As depicted in Figure 4(a), the reconstructed $H(z)$ using RBFNN is statistically aligned with the curve of the flat $\Lambda$CDM model, showing similar trends at low redshifts, although revealing a notable discrepancy at higher redshifts when compared to the lower redshifts. This anomalous phenomenon is caused by the lack of covariance in the observational data.

### 4. Reconstructing OHD Data Using Covariance

To verify the hypothesis that the lack of covariance may lead to inaccurate prediction results as mentioned in the previous section, we propose a new method for reconstructing OHD data using redshift pairs and their corresponding covariance in this section. In Section 4.1, we describe a redshift-pair reconstruction method based on the Cartesian product, in Section 4.2, we introduce the process of reconstructing OHD data using observational covariance, and in Section 4.3, we present the validation and analysis of the results.

#### 4.1. Redshift-pair Reconstruction Method Based on the Cartesian Product

In this section, we propose a new method for reconstructing OHD data based on the Cartesian product of redshift pairs. The Cartesian product is defined on an ordered set and is the set of all possible ordered combinations of elements in the sets, each element of which is included in these sets. Specifically, we reconstruct the input of redshift point $z$ as a redshift pair $(z_1, z_2)$ and reconstruct the output $(H(z), \sigma(z))$ as $(H(z_1), \sigma(z_1),$ and $H(z_2), \sigma(z_2)$. There are 32 observed values, and each data is composed of $(z, (H(z), \sigma(Hz))$. For example, $(z_1, H(z_1), \sigma(z_1))$, $(z_2, H(z_2), \sigma(z_2))$, and $(z_3, H(z_3), \sigma(z_3))$ are the first three observed data. After transformation, the data are represented as $[z_i, z_j, H(z_i), \sigma(z_i),$ and $H(z_j), \sigma(z_j))]$, where $i, j \leqslant 32$. Keeping the monotonicity of the Hubble parameter with respect to the redshift helps to reconstruct the monotonicity of the data better, and in subsequent studies, the covariance between the observed points and the Hubble parameter is required. Therefore, we first perform a Cartesian product instead of a permutation or combination on the set of 32 observed redshift points, which helps to preserve the monotonicity of the Hubble parameter. As a result, we obtain a set of 1024 (i.e., $32 \times 32$) elements, which is done to study the effect of the covariance between the Hubble parameter values. In this process, a $1024 \times 2$ matrix ($z_i$, $z_j$) is fed into an RBFNN and trained with the optimal hyperparameters determined in Section 3.2, which outputs a $1024 \times 4$ matrix ($H(z_i)$, $\sigma_{(z_i)}$, $H(z_j)$, $\sigma_{(z_j)}$). Because for each predicted point, multiple $H(z)$ and $\sigma_{(z)}$ are obtained using the above method, such as when 1024 redshift pairs are input, each redshift point will obtain 64 (32+32) corresponding $H(z)$ and $\sigma_{(z)}$. Therefore, it is necessary to extract all $H(z)$ and $\sigma_{(z)}$ for each redshift point $z$ and finally estimate a unique $H(z)$ and $\sigma_{(z)}$ for each redshift point $z$ by taking their mean. The reconstructed result is shown as the orange curve in Figure 5.

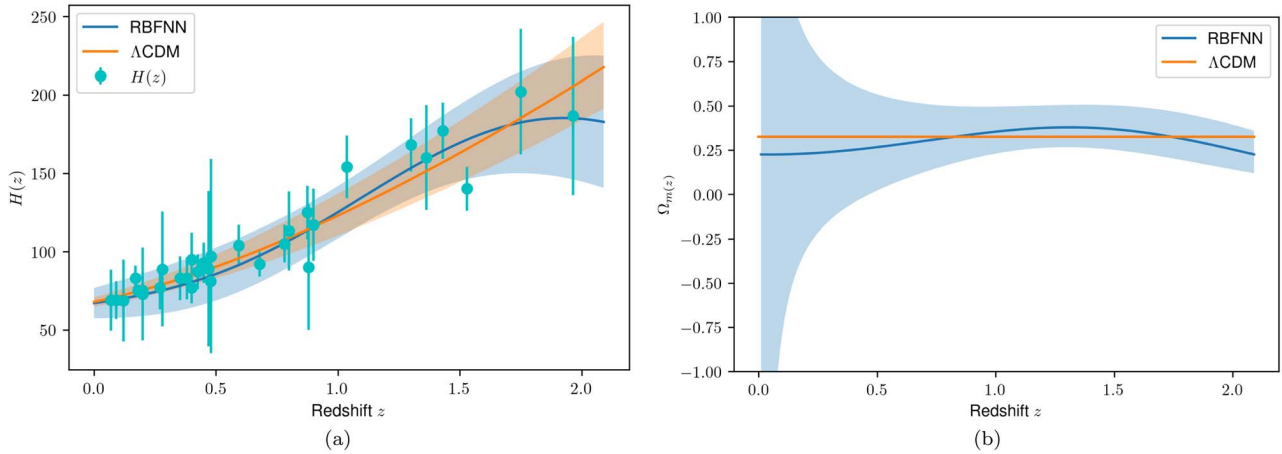

**Figure 4.** The reconstructed functions of observational $H(z)$ (solid blue curve) and its corresponding $1\sigma$ error with RBF networks (a). Comparison between the reconstructed $\Omega_m$ values using the RBF network based on simulated OHD data and the $\Lambda$CDM ($\Omega_m = 0.326$)(b). In panel (a), the green dots with error bars represent the simulated $H(z)$ data and the corresponding $1\sigma$ error, while the orange curve corresponds to the fiducial flat $\Lambda$CDM model with $H_0 = 67.824$ km s$^{-1}$ Mpc$^{-1}$ and $\Omega_m = 0.326$. In the left panel, the shaded regions illustrate the reconstruction errors associated with both the $\Lambda$CDM model and the RBFNN model when using observational $H(z)$ data, while the right panel portrays the propagation error.

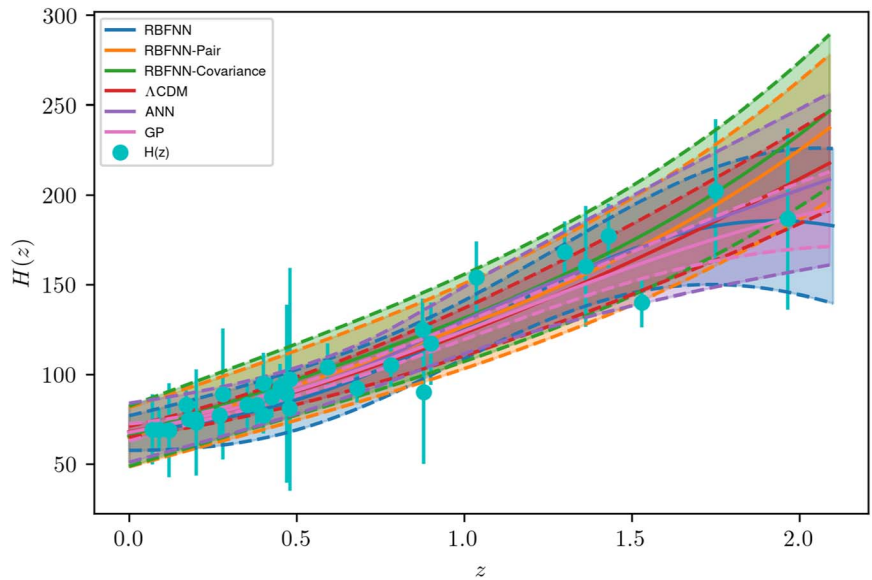

**Figure 5.** Comparison of the results from reconstructing OHD data using different methods.

### 4.2. Applying Observational Covariance to Reconstruct OHD Data

Based on the work in the previous section, we incorporate the covariance between the redshifts to further improve the reconstruction of the precise value of the Hubble constant.

Currently, the 31 available OHD data obtained from the cosmic chronometer method are derived from multiple galaxy samples and fitted with different methods. We believe that the data obtained by different methods are independent, and there is no covariance. Among them, 15 data were measured by Moresco et al. (2020) using the D4000 method. They systematically estimated various sources of errors in these data (statistical errors, contamination from young galaxies in the sample, galaxy spectral models, stellar formation history, etc.) and provided a method for calculating the covariance matrix. In this paper, we calculated the covariance of the relevant data based on their open-source program.[9] The calculation steps are as follows: Because the systematic bias caused by the choice of the star formation history and the age bias is covered by the data error, only the bias caused by the selection of the galaxy spectral model (distinguished by $a$, $b$) needs to be calculated.

---

[9] https://gitlab.com/mmoresco/CCcovariance/-/tree/master

(1) First, a synthetic simple stellar population (SSP) spectral library is generated based on a single stellar population synthesis model, which fully considers the impact of various attributes (SSP model, stellar age and metallicity, initial mass function, stellar library, etc.) on galaxy spectral modeling. Each different SSP model is marked with a letter, for example, $a$ or $b$.
(2) Measure the $D4000$ of all spectra in the library and construct the $D4000$-age relation based on different models.
(3) Use this relation to generate $\Delta z/D4000$ indicators for each model in the library using simulated $D4000(z)$ data (without noise).
(4) Use piecewise linear fitting to estimate the slope $A(Z)$ of the $D4000$-age relation calculated in (2), which is obtained by fitting a linear function to different $D4000$ ranges. For each model, this method can obtain several slopes $A$ as functions of the $D4000$ range.
(5) Estimate the Hubble parameter $H(z)$ by extracting the differential $D4000$ from simulated data generated from model $a$ and the slope obtained from model $b$, and construct the $H(z)$ deviation matrix percentage $\eta(z)_{ab}$ for combinations of different models.
(6) Finally, the $H(z)$ deviation matrix percentage $\eta(z)_{ab}$ is estimated using the following formula and propagated to the average deviation percentage $\hat{\eta}(z)$:

$$\eta(z)_{a,b} = \frac{H(z)_{a,b} - H_{\mathrm{fid}}(z)}{H_{\mathrm{fid}}(z)}. \quad (15)$$

The covariance, $\mathrm{Cov}_{i,j}^X$, between any two redshift points in OHD data is defined in this article as follows:

$$\mathrm{Cov}_{i,j}^X = \hat{\eta}^X(z_i) \times H(z_i) \times \hat{\eta}^X(z_j) \times H(z_j). \quad (16)$$

The superscript X denotes the source of the error.

In summary, the first step is to apply the redshift-reconstruction method based on the Cartesian product proposed in this article and input the 961 × 2 matrix $(z_i, z_j)$ into an RBFNN. Unlike the redshift-reconstruction method based on the Cartesian product, an additional fifth column is added in the output layer, which represents the observation covariance between redshift pairs. This results in an output matrix of 961 × 5 denoted as $(H(z_i), \sigma(z_i), \mathrm{Cov}_{i,j})$ (as verified by experiments in this article, separately fitting a univariate model and directly fitting a multivariate model have a significant impact on the results).

### 4.3. Results and Discussion

This article uses the chi-square test to analyze the deviation between the reconstruction results and observed OHD data in order to verify the hypothesis that the addition of covariance can effectively reduce reconstruction errors. The chi-square test, also known as the goodness-of-fit test, is a nonparametric test. Because nonparametric tests do not assume specific parameters or a normal distribution of the population, they are sometimes referred to as distribution-free tests. This statistical method allows us to compare the observed data $O$ to the expected values $E$ and to determine the extent of their deviation

**Table 2**
Comparison of the Chi-square Test Results

| Model    | Sum  | Mean | Median |
|----------|------|------|--------|
| GP       | 12.4 | 0.40 | 0.08   |
| RBFNN(B) | 5.55 | 0.18 | 0.06   |
| RBFNN(A) | 5.48 | 0.18 | 0.05   |
| ANN      | 4.34 | 0.14 | 0.06   |
| RBFNN(C) | 4.17 | 0.14 | 0.03   |

in the form of

$$\chi^2 = \sum \frac{O - E}{E}. \quad (17)$$

This will enable us to determine which of the methods is most reliable for reconstructing data, and it will provide a basis for making informed decisions about which method to use. Combined with the application of the chi-square formula in cosmology, the chi-square value in this article is given by the following formula:

$$\chi^2 = \sum \frac{(H_p - H_o)^2}{\sigma_{H_p}^2 + \sigma_{H_o}^2}, \quad (18)$$

where $H_p$ represents the predicted $H(z)$ corresponding to the redshift $z$ based on the OHD data, $H_o$ represents the observed $H(z)$ corresponding to the redshift $z$ in the OHD data, $\sigma_{H_p}$ represents the predicted $\sigma(z)$ corresponding to the redshift $z$ in the OHD data, and $\sigma_{H_o}$ represents the observed $\sigma(z)$ corresponding to the redshift $z$ in the OHD data. We perform a statistical analysis on a data set consisting of 31 data points extracted from the OHD, and the results are summarized in Table 2. Specifically, for the GP model, we calculate a chi-square sum of 12.40, a mean of 0.40, and a median of 0.08. Notably, this chi-square score surpasses the scores obtained from the four neural network-based models by a significant margin. Furthermore, we compute the chi-square sum, mean, and median for the OHD data reconstructed using the redshift-pair method based on the Cartesian product. These values amount to 5.55, 0.18, and 0.06, respectively, which are slightly lower than the values achieved by the OHD data reconstructed using the RBFNN. Both of these values are also slightly lower than those obtained using the ANN method, which we replicated based on the work of Wang et al. (2020). However, it is worth mentioning that when observation covariance is taken into account, all of the above results are surpassed by the redshift-pair reconstruction method based on the Cartesian product. Specifically, this method yields a chi-square sum of 4.17, a chi-square mean of 0.14, and a chi-square median of 0.03. These results clearly demonstrate the superiority of the redshift-pair reconstruction method based on the Cartesian product when observation covariance is included in the analysis.

We visualize the reconstruction results of the three proposed models in this article and compare them with the ANN, GP, and $\Lambda$CDM model, as shown in Figure 5. In the figure, RBFNN represents the curve reconstructed by applying RBFNN to OHD data, RBFNN-pair represents the curve reconstructed by applying the redshift pairs based on the Cartesian product method to OHD data, RBFNN-covariance represents the curve reconstructed by incorporating observation covariance into OHD data reconstruction, and ΛCDM represents the curve fitted by using the ΛCDM model to reconstruct OHD. As can be seen from Figure 5, the redshift pairs based on the Cartesian product method overcome the limitation of independent data points in the original RBFNN model (RBFNN(A)) and show an upward trend at high redshifts.

## 5. Conclusion and Discussions

In this study, we use a model-independent approach based on ANNs to reconstruct $H(z)$ data and obtain a reliable value of $H_0$. We first simulate 31 available observational $H(z)$ data based on a flat ΛCDM model with fiducial values of $H_0 = 70 \text{ km s}^{-1} \text{ Mpc}^{-1}$ and matter energy density parameter $\Omega_m = 0.3$. We then use these simulated data points to reconstruct the function of $H(z)$ using an RBFNN model and perform hyperparameter tuning using a fivefold GS cross validation to obtain the optimal hyperparameters for the RBFNN model. Next, we train the RBFNN model using the observed $H(z)$ data with the optimal hyperparameters, obtaining a final reconstruction result of $H_0 = 67.1 \pm 9.7 \text{ km s}^{-1} \text{ Mpc}^{-1}$. Our study highlights the capability of a model-independent method to reconstruct $H(z)$ using ANN models, which can be more reliable than the GP methods. We compared this result with the fiducial flat model with $H_0 = 67.824 \text{ km s}^{-1} \text{ Mpc}^{-1}$ and $\Omega_m = 0.326$ and found that the two curves had similar trends, with a relatively noticeable downturn in the RBFNN curve at high redshifts. This is because when using RBFNNs to reconstruct data, the assumption is made that the data points are independent of each other, so that the reconstruction result at high redshifts is more influenced by a small number of outlier data points, resulting in a deviation from the theoretical prediction. Based on extensive research within the academic community on the expansion history of the Universe, it has been shown that the observed downward trend in reconstructing this history is most likely due to a lack of data points at high redshifts, where they are often outliers. This presents a significant obstacle for neural network models to accurately reconstruct the expansion history, and it consequently leads to an increase in errors. Furthermore, because the expansion history of the Universe is a relatively smooth curve, neural network models are susceptible to overfitting or underfitting issues that may lead to inaccurate predictions. Therefore, when using neural network models to reconstruct the expansion history of the Universe, it is essential to consider other factors comprehensively and use alternative methods for verification and adjustment to ensure the accuracy and reliability of the results. To mitigate these problems, we suggest increasing the covariance of the data, which can introduce more relevant information and reduce the sensitivity to outliers.

To verify our hypothesis and address the challenge of insufficient data points at high redshifts, we propose a redshift-pair-based reconstruction method established on the Cartesian product. This method lays the foundation for using covariance data, which can introduce more relevant information and increase the correlation between data points. We first test the reliability of this method by performing chi-squared tests, which show that it has a significant impact on the $H(z)$ reconstructing process. We further increase the correlation between data points by incorporating observational covariance between redshifts, and once again reconstruct the OHD data using RBFNNs, while observing the impact of the addition of covariance on the reconstruction results. After testing, the RBFNN model with observed covariance performed better in the chi-squared test, significantly improving the reliability of the RBFNN reconstruction of the OHD data. Therefore, it can be concluded that the RBFNN redshift-pair-based method can effectively reconstruct the overall information of $H(z)$, and adding data covariance can further increase the correlation between data points. After the chi-squared test, we validate the reconstructed results of our new method by comparing them with the observed data. The comparison shows that the reconstructed results of our new method are more consistent with the observed data, indicating that the incorporation of covariance into the modeling process can effectively address the challenges posed by outliers and overfitting or underfitting issues.

In conclusion, this study highlights the effectiveness of a model-independent approach based on ANNs in reconstructing $H(z)$ data and expands its application to make it more reliable. Additionally, we propose a redshift-pair-based reconstruction method that laid the foundation for using covariance data. Looking forward, there are several potential areas for future research that could further enhance the reliability and accuracy of OHD reconstruction. Exploring alternative neural network architectures, such as convolutional neural networks or long short-term memory networks, would help reconstruct the complex features in the expansion history of the Universe. Another area involves improving the quality and quantity of observational data, which would significantly impact the accuracy and reliability of the reconstruction methods. By further investigating these areas, we can gain a deeper understanding of the expansion history of the Universe, its properties, and the role of neural network-based methods in cosmology, enabling us to make more precise and robust predictions of the behavior of the Universe.

<>
## Acknowledgments

We are grateful for the insightful and useful comments of the referee that helped us improve our manuscript. T.J.Z. (张同杰) dedicates this paper to the memory of his mother, Yu-Zhen Han (韩玉珍), who passed away 3 yr ago (2020 August 26). We are grateful to Zhen-Zhao Tao, Jing Niu, and Yu-Chen Wang for their useful discussions. This work was supported by the National Science Foundation of China (grants No. 11929301).

## ORCID iDs

Jian-Chen Zhang  https://orcid.org/0000-0002-1428-8311
Yu Hu  https://orcid.org/0009-0002-1805-4288
Kang Jiao  https://orcid.org/0000-0003-0167-9345
Tong-Jie Zhang
 (张同杰)  https://orcid.org/0000-0002-3363-9965

## References

Bernardo, R. C., & Said, J. L. 2021, JCAP, 2021, 027
Broomhead, D. S., & Lowe, D. 1988, Radial Basis Functions, Multivariable Functional Interpolation and Adaptive Networks 4148, Royal Signals and Radar Establishment Malvern https://apps.dtic.mil/sti/citations/ADA196234
Chen, J. F., Wang, Y. C., Zhang, T., & Zhang, T. J. 2023, PhRvD, 107, 063517
Colgáin, E. Ó., & Sheikh-Jabbari, M. M. 2021, arXiv:2101.08565
Dhanalakshmi, P., Palanivel, S., & Ramalingam, V. 2009, Expert Syst. Appl., 36, 6069